\documentclass[12pt]{article}

\usepackage[margin=1in]{geometry}
\usepackage{graphicx}
\usepackage{epstopdf}
\usepackage{caption}
\usepackage{float}
\usepackage{amssymb,amsmath}
\usepackage{color}
\usepackage{pstricks}
\usepackage{pifont}
\usepackage[round,comma,authoryear]{natbib}

\usepackage{comment}

\usepackage[]{hyperref}
\hypersetup{colorlinks,citecolor=blue}
\usepackage[capitalise]{cleveref}

\usepackage{longtable}
\usepackage[title]{appendix}

\def\BIBand{and}

\def\nvec#1{\mathbf {#1}}

\def\nten#1{\mbox{\boldmath{$\rm #1$}}}

\def\St{\partial\Omega_{\rm t}}
\def\Su{\partial\Omega_{\rm u}}

\title{A discrete dislocation analysis of size-dependent plasticity in torsion}
\author{A. Cruzado$^1$, M.~P. Ariza$^2$, A. Needleman$^3$, M. Ortiz$^4$, A.~A. Benzerga$^{1,3}$ \\
\footnotesize $^1$Department of Aerospace Engineering, Texas A\&M University, College Station, TX 77843\\
\footnotesize $^2$Escuela T\'ecnica Superior de Ingenier\'ia, Universidad de Sevilla, Sevilla 41092, Spain\\
\footnotesize $^3$Department of Materials Science $\&$ Engineering, Texas A\&M University, College Station, TX 77843\\
\footnotesize $^4$Division of Engineering and Applied Science, California Institute of Technology, Pasadena, CA 91125}

\begin{document}
\maketitle

\thispagestyle{empty}

\begin{abstract}
A method for solving three dimensional discrete dislocation plasticity
boundary-value problems
using a monopole  representation of the dislocations is presented.  At each
time step, the displacement, strain and stress fields in a finite body are
obtained by superposition of  infinite body dislocation fields 
and an image field that enforces the boundary
conditions. The three dimensional  infinite body fields are obtained
by representing  dislocations as being comprised of points, termed
monopoles, that carry 
dislocation line and Burgers vector information.  
The image fields are obtained from a three dimensional linear
elastic finite element 
calculation.  The implementation of the coupling of the monopole
representation with the finite element method, including the interaction of
curved dislocations with  free
surfaces,  is presented in some
detail because it differs significantly from  an implementation
with a line based dislocation representation.  
Numerical convergence and 
the modeling of dislocation loop nucleation
for large scale  computations are  investigated. 
The monopole discrete dislocation plasticity framework is used to
investigate the effect of size and 
initial dislocation density on the torsion of wires with diameters
varying over three orders of magnitude. Depending on the initial
dislocation source density and the wire diameter, three regimes of 
torsion-twist response are obtained:
(i) for wires with a sufficiently small diameter,  plastic deformation
is nucleation controlled and is strongly size dependent; (ii) for
wires with larger diameters 
dislocation plasticity  is dislocation   interaction controlled, with
the emergence of  geometrically necessary dislocations and dislocation
  pile-ups playing a key role, and is strongly size dependent; and
  (iii) for wires with sufficiently 
  large diameters plastic deformation becomes less heterogeneous and
  the dependence on  size is greatly diminished.
\end{abstract}


\section{Introduction}

Plastic deformation arising from the evolution of dislocations is
size dependent in micron scale metal crystals, see e.g. 
\citet{Hutchinson00}. There are a variety of mechanisms that can lead
to such size dependence with smaller being stronger. Some, for
example, dislocation starvation and single arm dislocation source
controlled plasticity at surfaces, can lead to size dependence when
the imposed boundary conditions are consistent with homogeneous
deformation \citep{Uchic09,Awady15}.  When the imposed boundary
conditions and/or boundary constraints  induce a deformation gradient,
dislocation structures evolve that have a local net Burgers vector,
the so-called Geometrically Necessary Dislocations (GNDs)
\citep{Ashby70}, give rise to long range stresses and induce a strong
size effect. Examples of loadings and  boundary constraints that lead
to a strong size effect include indentation \citep{Nix98, Kysar07},
beam bending \citep{Stolken98,Cleveringa99}, torsion
\citep{Fleck94}, passivated thin films \citep{Nicola06}, and in the
vicinity of a crack tip or sharp notch \citep{Cleveringa00}.

Dislocations in crystals are line defects. Hence, in mesoscale
modeling of the evolution of dislocations in crystals, the
dislocations are modeled as line defects in an elastic solid. In three
dimensions, tracking the evolution of such lines is computationally
complex \citep{Bulatov98,Zbib98,Ghoniem00,Arsenlis07}. More recently,
\citet{Deffo19} have introduced an alternative description where the
dislocation lines are carried by points, termed monopoles, and it is
these points that must be kept track of, rather than the lines they
carry. This has the potential for significantly simplifying mesoscale
dislocation plasticity modeling.
  
In this study, torsion of single crystal wires is modeled using the
monopole dislocation representation of \cite{Deffo19, ARIZA21}. The
method of monopoles may be regarded as a particle method applied to
line dislocations, in which the dislocation line is concentrated at
points, while simultaneously retaining local geometric information
about the lines, namely length and direction, as well as local Burgers
vector information. The collection of dislocation monopoles then
supplies an approximation, in the sense of local averages, to the
dislocation density, in the spirit of particle methods for mass
transport \citep{Fedeli:2017, carrillo2017, carrillo2019,
  Pandolfi:2023}. The dislocation monopoles are updated in time
according to linear-elastic energetic driving forces, a mobility law
and geometric transformations that take into account the stretching of
the dislocation lines due to curvature. These geometric updates encode
the requirement that the monopoles approximate lines and, in
particular, they ensure that the zero-divergence property of the
dislocation density is satisfied approximately at all times. The
method of monopoles effectively eschews the problems of line
entanglement that inevitably arise in line-based representations and,
as is commonly the case with particle methods, is remarkable for its
ability to navigate through complex three-dimensional dislocation
dynamics simply and without {\sl ad hoc} rules.

Three dimensional quasi-static calculations are carried out, within
the context of a small deformation formulation (geometry changes
neglected), for wires with diameters ranging from 150nm to
10000nm. Various initial dislocation source densities 
are considered. Boundary conditions are imposed using the
superposition method introduced by \citet{Giessen95} and, for line
representation of dislocations, extended to three dimensions by
\citet{Weygand02,Crone14,Joa23}; also see \citep{Vattre14,Bertin15}
for alternative methods. The monopole representation of dislocations
requires a different treatment of dislocation-free boundary
interactions than that given by \citet{Weygand02} or more recent
line-based methods. 

Experimentally, a strong size effect in torsion of polycrystalline metal
wires was documented by \citet{Fleck94}. Because wire torsion is an 
inherently three dimensional configuration, discrete dislocation modeling of 
wire torsion has only been carried out relatively recently by 
\citet{Senger11}, \citet{GRAVELL2020}, and \citet{RYU20}.
These analyses were carried out for single crystals and used a line 
representation of the dislocations. As in the experiments of \citet{Fleck94}, 
a strong size effect was obtained.

The calculations here are carried out for wires with a circular cross
section comprised of a single fcc crystal with properties
representative of copper. Because implementation of the coupling
between the monopole dislocation representation and the finite element
method differs significantly from that for the line representation
finite element coupling, the implementation of the coupling is
described in some detail. Other aspects of the numerical
implementation are discussed including consideration of numerical
accuracy. Results are then presented for the effect of size on the
moment-twist relation and for the effect of size on the dislocation
structures that emerge. The variation of the scaling of the
moment-twist relation with initial dislocation density is also
illustrated. 

\section{Discrete dislocation plasticity framework}

The three dimensional discrete dislocation plasticity formulation is based on the monopole representation of dislocations introduced by \citet{Deffo19}, also see \citet{ARIZA21}, together with the superposition of an image field following \citet{Giessen95} and \citet{Weygand02}. The presentation of the framework here focuses on the aspects particular to combining the monopole dislocation representation with the image field obtained from a finite element calculation.

\begin{figure}[h]
   \centering
    \includegraphics[width=\textwidth]{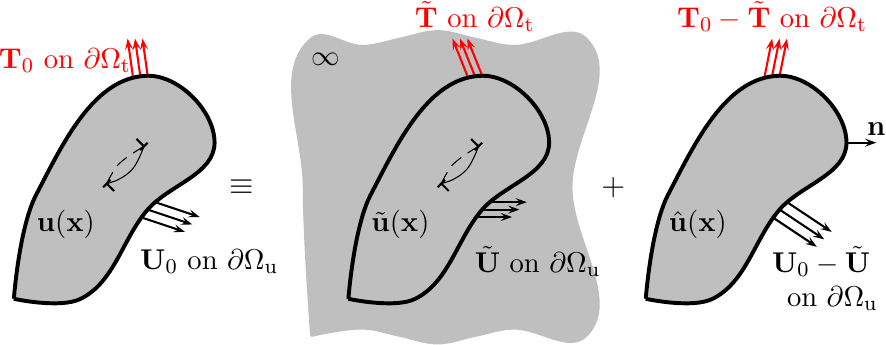} 
\caption{Decomposition of the problem into the superposition of
  interacting dislocations in a homogeneous infinite solid, the (
  $\tilde{}$ ) fields, and the  image problem that enforces the
  boundary conditions, 
   the ( $\hat{}$ )   fields. } 
\label{fig:superpos} 
\end{figure}

A finite body occupying domain $\Omega$ is subjected to boundary tractions 
$\nvec T_0$ on $\partial\Omega_{\rm t}$ and boundary displacements 
$\nvec U_0$ on $\partial\Omega_{\rm u}$, as sketched in \cref{fig:superpos}. 
In its current state, the body contains a distribution of dislocation lines, 
collectively denoted $C$, carrying a Burgers vector $\nvec{b}(\nvec x)$ 
at every point $\nvec x$ of $C$. The dislocation network must be closed, 
i.~e., the corresponding Nye's dislocation density \citep{nye:1953} 
must be divergence free, or terminate at the boundary of the solid. 
In addition, the Burgers vector $\nvec{b}(\nvec x)$ must be conserved 
along dislocation lines and satisfy Frank's rule at branching points. 

The procedure used to calculate the evolution in time of the dislocation
configuration for given time-dependent  boundary tractions
and boundary displacements  is
described in the following sections.  

\subsection{The monopole approximation}

In calculations, we approximate the dislocation distribution by an ensemble of $M$ monopoles labeled $a = 1, \dots, M$, with coordinates $\{\nvec{x}_a\}_{a=1}^{M}$. The monopoles additionally carry corresponding elements of dislocation line $\{\boldsymbol{\xi}_a\}_{a=1}^{M}$ and Burgers vector $\{\nvec{b}_a\}_{a=1}^{M}$. The monopole ensemble approximates the dislocation line density in the sense that, for every continuous function $\varphi_{ij}(\nvec x)$, 
\begin{equation} \label{1QMIwo}
    \oint_C \varphi_{ij}(\nvec x) b_i(\nvec x) dx_j
    \sim
    \sum_{a=1}^M \varphi_{ij}(\nvec{x}_a) b_{a i} \xi_{a j} ,
\end{equation}
where we write $b_{a i} := (\nvec{b}_a)_i$ and $\xi_{a i} := (\boldsymbol{\xi}_a)_i$, and the error of the approximation can be controlled and made arbitrarily small by introducing a sufficiently large number of monopoles. 

By Stokes$'$ theorem, the divergence, or closedness condition of the
dislocation line density, requires that 
\begin{equation}
    \oint_C \frac{\partial f_i}{\partial x_j} (\nvec x) b_i(\nvec x) dx_j
    =
    0 ,
\end{equation}
for every differentiable function $f_i(\nvec x)$ with finite support. 
By the assumed approximation property, Eq.~(\ref{1QMIwo}), it follows that, 
for a distribution of monopoles to represent an admissible distribution of closed dislocation lines, it must satisfy
\begin{equation} \label{QGF8wP}
    \sum_{a=1}^M \frac{\partial f_i}{\partial x_j} (\nvec{x}_a) b_{a i} \xi_{a j} 
    \sim
    0 
\end{equation}
for every differentiable function $f_i(\nvec x)$ with finite support, to within the level of approximation of the monopoles. This condition places a geometrical constraint on the distribution of monopoles that must be enforced at all times during the calculation by means of suitable geometric updates, cf.~Section~\ref{AvkokR}.

\subsection{The instantaneous boundary-value problem}

Suppose that the body is in equilibrium and the position of every monopole 
in the body is known. Superposition may then be used to write the displacement, ${\nvec u}$, strain, $\nten \varepsilon$, and stress, $\boldsymbol{\sigma}$, fields as
\begin{equation}
   {\nvec u} (\nvec x) = \tilde{\nvec u}(\nvec x) + \hat{\nvec u}(\nvec x)
   \,; \qquad
   \nten \varepsilon = \tilde{\nten\varepsilon} + \hat{\nten\varepsilon} 
   \,; \qquad
   \nten\sigma = \tilde{\nten\sigma}  + \hat{\nten\sigma} \,,
\label{e:sup}
\end{equation}
where the symbol $\tilde{(\cdot)}$ refers to infinite-body fields and $\hat{(\cdot)}$ 
to image fields that enforce the boundary conditions, \cref{fig:superpos}.
By linearity, the infinite-body displacement and stress fields follow as
\begin{equation} 
    \tilde{\nvec u}(\nvec x) = \sum_{a=1}^M  {\nvec u}_a (\nvec x) \,,
    \qquad
	\tilde{\nten \sigma}(\nvec x) = \sum_{a=1}^M  \nten \sigma_a (\nvec x) \,,
 \label{e:sigma_tilde}
\end{equation}
where ${\nvec u}_a (\nvec x)$ and $\nten \sigma_a (\nvec x)$ are the contributions to the fields from monopole $a$, respectively. The corresponding expressions follow by inserting the approximation property Eq.~(\ref{1QMIwo}) into well-known integral expressions from linear-elastic dislocation theory, cf., e.~g., \citep[Chapter 4]{HirthBook82},
and are collected in the appendix, Eqs.~(\ref{hQaA0W}) and (\ref{IuizhN}), for completeness.

The $\hat{(\cdot)}$ fields are obtained from the solution of the
linear elastic boundary value problem specified by 
\begin{equation}
            \nabla \cdot \hat{\nten\sigma} = \nvec 0 
   \quad ; \quad
   \hat{\nten\varepsilon} = \nabla \otimes \hat{\nvec u}
   \quad ; \quad
   \hat{\nten\sigma} = \nten{\cal L} : \hat{\nten\varepsilon} 
   \qquad \mbox{for} \, \nvec x \in \Omega,
  \label{e:BVP}
\end{equation}
\begin{equation}
\nvec n \cdot \hat{\nten\sigma} 
   = \nvec T_0 - \nvec n \cdot \tilde{\nten\sigma}
   \quad \mbox{for} \, \nvec x \in \St
   \quad ; \quad
   \hat{\nvec u} = \nvec U_0 - \tilde{\nvec U}
   \quad \mbox{for} \, \nvec x \in \Su
\label{e:BCs}
\end{equation}
with $\nten{\cal L}$ the isotropic tensor of elastic moduli  
and $\nvec n$ the outer normal to $\partial\Omega$. 
Also, $\tilde{\nvec U}$ and $\tilde{\nvec T} = \nvec n \cdot \tilde{\nten\sigma}$
are respectively the boundary displacement and traction arising from the infinite body
dislocation fields (see \cref{fig:superpos}),
$\nabla$ is the nabla operator and $\otimes$ stands for the symmetric dyad.
This boundary value problem is solved by the finite element method and traction boundary conditions are imposed through effective nodal forces. With ${\nvec F}_{\rm 0,n}$ denoting the imposed nodal forces on a set $\Gamma_{\mathrm{f}}$ of boundary nodes lying on $\St$, and with $\hat{\nvec f}_{\rm n}$ and $\tilde{\nvec f}_{\rm n}$ the corresponding nodal forces associated with the $\hat{(\cdot)}$ and $\tilde{(\cdot)}$ fields, the nodal forces imposed to obtain the image fields are 
\begin{equation}
 \hat{\nvec f}_{\rm n}= {\nvec F}_{\rm 0,n}-  \tilde{\nvec f}_{\rm n},
 \qquad {\rm n} \in  \Gamma_{\mathrm{f}}  
\label{e:trac_hat}
\end{equation}
The values of the tilde forces $\tilde{\nvec f}_{\rm n}$ are obtained by assembling element nodal forces from all elements that have a boundary surface node. For each such element its contribution to the surface boundary nodal force is obtained by numerical integration of $\tilde{\nten\sigma}(\nvec x)$ within that element.

\subsection{Geometric updates} \label{AvkokR}

Suppose that the velocities $\{\nvec{v}_a\}_{a=1}^M$ of the monopoles are known. Then, the positions $\{\nvec{x}_a\}_{a=1}^M$ of the monopoles can be updated simply by time integration. However, in order to fully update the monopole configuration we additionally need update rules for the Burgers vectors $\{\nvec{b}_a\}_{a=1}^M$ and line elements 
$\{\boldsymbol{\xi}_a\}_{a=1}^M$. The update rules follow by time integration of the transport, or continuity, equation for the dislocation lines (cf.~\cite{Mura82}, Eqs.~(32.2) and (38.4)). 
For a distribution of monopoles, by the approximation property, \cref{1QMIwo},
the discrete form of the transport equations becomes \citep{Deffo19, ARIZA21}
\begin{equation} \label{0LetiE}
    \dot{\nvec{b}}_{a} = \nvec{0} ,
    \qquad
    \dot{\boldsymbol{\xi}}_{a}
    =
    (\nabla\nvec{v}_a) \boldsymbol{\xi}_{a} ,
\end{equation}
where $\nabla\nvec{v}_a$ is the gradient at $\nvec{x}_a$ of some suitable spatial interpolation of the monopole velocities $\{\nvec{v}_a\}_{a=1}^M$ and $(\nabla\nvec{v}_a) \boldsymbol{\xi}_{a}$ is the gradient of the velocity field in the direction of the line element $\boldsymbol{\xi}_{a}$. 

It can be readily verified using the approximation property
in Eq.~(\ref{1QMIwo}) that the update of Eq.~(\ref{0LetiE}) properly
accounts for the stretching and rotation of the line elements
attendant to the motion of the monopoles \citep{ARIZA21}. In addition,
it can also be verified \citep{Deffo19} that the update in
Eq.~(\ref{0LetiE}) preserves dislocation loops remaining closed in the sense of
Eq.~(\ref{QGF8wP}). Thus, if the monopole ensemble initially
approximates a closed distribution of dislocation lines, it still does
so after the update, to within the accuracy of the monopole
approximation and the velocity interpolation.  

The computation of the velocity gradient requires an interpolation scheme of the general form
\begin{equation}\label{x5Hus6}
    (\nabla \nvec{v}_a)_{ij}
    =
    \sum_{b=1}^M v_{b i} \frac{\partial N_{b}}{\partial x_j}(\nvec{x}_a) .
\end{equation}
where $\{N_a(\nvec x)\}_{a=1}^M$ are suitable shape functions. In calculations we specifically use the max-ent shape functions \citep{ArroyoOrtiz:2006}
\begin{equation}\label{b4lEfR}
    N_{a}(\nvec{x})
    =
    \frac{1}{Z(\nvec{x})}
    \exp\left(-\frac{\beta_a}{2} |\nvec{x}-\nvec{x}_{a}|^2\right) ,
    \qquad
    Z(\nvec{x})
    =
    \sum_{a=1}^M
        \exp\left(-\frac{\beta_a}{2} |\nvec{x}-\nvec{x}_{a}|^2\right) .
\end{equation}
We note that the resulting interpolation, Eq.~(\ref{x5Hus6}), is
mesh-free and does not entail any ordering or connectivity between the
monopoles. In Eq.~(\ref{b4lEfR}), the distance $1/\sqrt{\beta_a}$ sets
the range of interaction for monopole $a$. Thus, only those monopoles
that are at a distance of order $1/\sqrt{\beta_a}$ `see' and interact
with monopole $a$. In practice, this property can be exploited to
reduce the range of the sums in Eq.~(\ref{x5Hus6}) and
Eq.~(\ref{b4lEfR}). 

The effect of the shape functions $N_{a}(\nvec{x})$ is to broaden the
monopoles and give them a spatial profile of finite width
$1/\sqrt{\beta_a}$. This broadening allows neighboring monopoles to
`see' each other for purposes of the interpolation of their velocities
into a continuous velocity field, which in turns governs the
geometrical update of the line elements
$\{\boldsymbol{\xi}_{a}\}_{a=1}^M$ per Eq.~(\ref{0LetiE}). In this
sense, the broadening of the monopoles is analogous to
finite-particle, or `blob', methods for problems of mass transport
\citep{carrillo2019, Pandolfi:2023}.  

\subsection{Mobility}

In order to close the evolution problem, a mobility law for the
velocity of the monopoles is required. The motion of the
monopoles is taken to be driven by the configurational Peach-Koehler
force  to account for a change in position of the dislocation as well
as the self force associated with the energy cost of a change in
dislocation line length, so that
\begin{equation}
    \nvec f_a 
    = 
    \big( {\nten\sigma}(\nvec x_a) \cdot {\nvec b_a} \big) \times {\boldsymbol{\xi}_a} ,
\label{e:PK}
\end{equation}
or, in view of Eq.~(\ref{e:sup}) and Eq.~(\ref{e:sigma_tilde}),
\begin{equation} \label{HzEtbK}
    \nvec f_a 
    = 
    ( \hat{\nten\sigma}(\nvec x_a) \cdot {\nvec b_a}) \times {\boldsymbol{\xi}_a}
	+ 
    \sum_{b=1}^M 
    ({\nten \sigma}_b(\nvec x_a) \cdot {\nvec b_a}) \times {\boldsymbol{\xi}_a} .
\end{equation}
The driving force $\nvec f_a$ comprises one term contributed by the
image stresses, which in turn accounts for boundary effects and the
finiteness of the domain, and a second term resulting from the elastic
interaction between the monopoles. 

However, a direct evaluation of the self-force terms,
$b=a$ in Eq.~(\ref{HzEtbK}), from linear elasticity is not possible
since ${\nten \sigma}_a(\nvec x)$ diverges as $\nvec x$ approaches
$\nvec{x}_a$. The singularity can be removed by fitting the monopoles
with a core profile of a width $\epsilon$ commensurate with a lattice parameter
\citep{BulatovCai:2006, Lazar:2017, Deffo19}. A particular core
profile, and the attendant regularization of monopole self-forces, can
be achieved by recourse to strain-gradient elasticity
\citep{Deffo19}. The resulting expressions are collected in 
\cref{app:monelas} for completeness.  

Once a suitably regularized set of monopole forces has been defined,
the velocity of the monopoles can be computed from a mobility law of
the type 
\begin{equation} \label{A0ae0C}
    \nvec{v}_a = \nten{M}_a \nvec{f}_a ,
\end{equation}
where $\nten{M}_a$ is a mobility tensor. In general, the edge
and screw mobilities of glide dislocations, as well as the climb mobilities,
differ greatly in actual
crystals, and this renders the mobility tensor $\nten{M}_a$
anisotropic. More general mobility laws, including nonlinearity, can
be treated likewise \citep{Deffo19, ARIZA21}. 

\subsection{Calculation of image forces}
Determining the first term in \cref{e:PK}, $\hat{\nvec{f}}_a$, requires evaluating
the image stress, determined by solving \cref{e:BVP,e:BCs}, at the
location of monopole $a$. 
This is accomplished by interpolating the $\hat{\nten \sigma}$ field
to the position of the monopole. During the deformation history, 
 monopoles change elements.
An efficient (${\cal O}(\log M)$ operations)
binary B-tree algorithm \citep{Comer79} (so-called BB-tree \citep{Zset20})
is used as an element locator to obtain the monopole
coordinates $\nvec {x}_{a}$ within an element.
Because  the $\hat{\nten \sigma}$ field is discontinuous across
element boundaries, the values of $\hat{\nten \sigma}$ are 
extrapolated to element nodes using the finite element shape
functions. Then the nodal values of $\hat{\nvec{f}}_a$
are averaged over all elements sharing the same node, before
interpolation to the monopoles.

\subsection{Topological transitions}

\begin{figure}[H]
   \centering
    \includegraphics[width=0.95\textwidth]{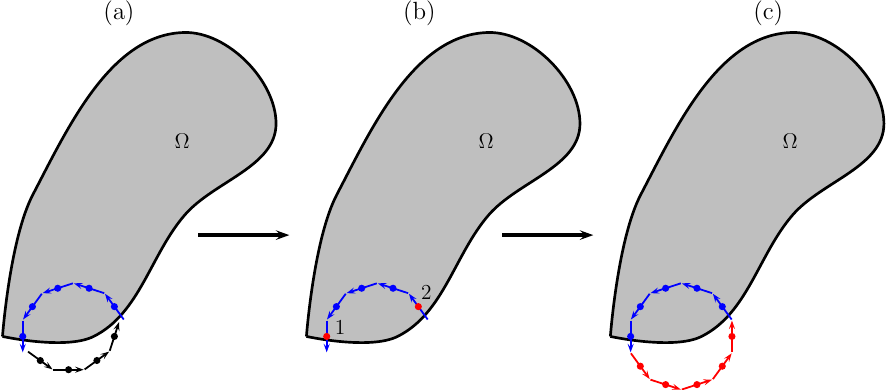} 
           
\caption{Schematic of a dislocation loop exiting the free surface and
  the introduction of virtual monopoles 
in three steps:
	(a) Step 1: check for monopoles that are outside the body
$\Omega$ (black) and remove them.  
	(b) Step 2: identify the two monopoles that are closest to the surface. 
	(c)  Step3: create virtual monopoles (red) and the corresponding element of
line $\xi$ that closes the loop.} 
\label{fig:virtual_loop} 
\end{figure}

Because the infinite body dislocation expressions pertain to a closed
line, a key issue in the numerical implementation of discrete dislocation
dynamics in finite bodies is the treatment of cases where part of a
dislocation loop exits the body,  as illustrated in
\cref{fig:virtual_loop}a. To address this within the context of the
monopole representation of dislocations, a method involving virtual
monopoles is introduced, akin to virtual segments employed  in
line-based methods \citep{Weygand02, Liu05, Weinberger09, Deng08,
  LEITER13, Crone14}. 

At each time step in the calculation, monopoles associated with a dislocation loop that have exited the body, as shown in \cref{fig:virtual_loop}a, are identified. Then the  two monopoles of that loop that remain in the body and that are closest to the surface are detected, see \cref{fig:virtual_loop}b. The distance between these two dislocations is denoted by $d_{1-2}$. Virtual monopoles are introduced to close the loop as illustrated in \cref{fig:virtual_loop}c.  The monopoles shown black in \cref{fig:virtual_loop}a are removed from subsequent calculations of the dislocation dynamics. The virtual monopoles shown in red in \cref{fig:virtual_loop}c are only used to calculate the tilde fields in \cref{e:PK,e:sigma_tilde}. In the computations here $10$ virtual monopoles lying on a semi-circle of radius equal to half the distance $d_{1-2}$ are added.

Dislocation nucleation/multiplication is included in the calculations 
via a source model \citep{Deffo19}. The locations of  potential sources are 
randomly distributed in the body prior to the start of the calculation 
of the deformation history.  Each potential source is defined 
as a set of monopoles arranged in a loop. 
At a given source location, a loop of radius $r_0$ nucleates when
\begin{equation}
    \alpha |\hat{\nvec{f}}_a| - |\mathbf{f}_a^{\epsilon}| > 0, 
    \quad 
    \forall a \in M(r_0),
    \label{eq:nuc}
\end{equation}
with $M(r_0)$ denoting the set of monopoles defining the loop and $\alpha > 0$ a parameter.
Here, $\mathbf{f}_a^{\epsilon}$ denotes the second term in \cref{e:PK},
regularized using a core parameter $\epsilon$; see \cref{app:monelas}.
\cref{eq:nuc} states that the external force acting on each monopole 
must exceed its internal force so that the loop does not collapse. 
Once this condition is met, the source emits a dislocation loop
so that the new monopoles that comprise the nucleated loop 
are inserted into the calculation.

\subsection{Coupled time-integration algorithm}

For given domain geometry and forcing, \cref{0LetiE,HzEtbK,A0ae0C}
govern the evolution of the monopole ensemble. 
Time integration may be accomplished using standard time-stepping algorithms, 
such as fully implicit backward-Euler \citep{Deffo19} 
or explicit two-stage Runge-Kutta method \citep{ARIZA21}.

The computational method involves the following steps:

\begin{enumerate}

\item At time $t$ the body 	is in equilibrium and the position of each monopole in the body is known. The surface nodal forces $\boldsymbol{\mathrm{\tilde{f}}}$ corresponding to 	$\boldsymbol{\tilde{\sigma}}$ and displacements $\boldsymbol{\tilde{u}}$ are computed.

\item The boundary-value problem defined by \cref{e:BVP,e:BCs} for the $\hat{(\cdot)}$ fields is solved. 

\item The $\boldsymbol{\hat{\sigma}}$ field is interpolated to the monopole locations 	to evaluate the Peach-Koehler forces $\{\nvec{f}_a\}_{a=1}^{M}$ in \cref{e:PK}. This interpolation is performed for both static source locations to evaluate the nucleation criterion, \cref{eq:nuc}, and for existing monopoles.

 \item The condition for new loop nucleation, \cref{eq:nuc}, is checked.

\item The rate equations Eq.~(\ref{0LetiE}), Eq.~(\ref{HzEtbK}) and Eq.~(\ref{A0ae0C}) 
	are integrated in time to determine new monopole configurations at time $t+\Delta t$.

\item Monopoles that lie outside the body are flagged and removed. Virtual monopoles are
	added to close the dislocation loop for stress calculations.

\item Go to step 1 with $t$ $\leftarrow$ $t+\Delta t$. 

\end{enumerate}

The code used to solve the initial/boundary value problem is a
combination of the serial MonoDis code of \citet{Deffo19,ARIZA21}  and the
Z-set \citep{Zset20}  finite element code. These are integrated
through a C++ plug-in interface. 
The implementation has the following features:
	(i) for the linear elastic finite element calculations, 
Cholesky decomposition of the global stiffness matrix is carried out once,
	(ii)   expressions derived from the principle of
virtual work are used to enforce the traction and displacement
boundary conditions, and 
	(iii) the image force interpolation uses a  BB-tree algorithm. 
Because  inversion of the global stiffness matrix is executed only
once, the time used for the 
finite element part solution is a small part of the total
computational time for calculation of the deformation history.

\section{Torsion problem formulation}

\begin{figure}[h!]
   \centering
    \includegraphics[width=\textwidth]{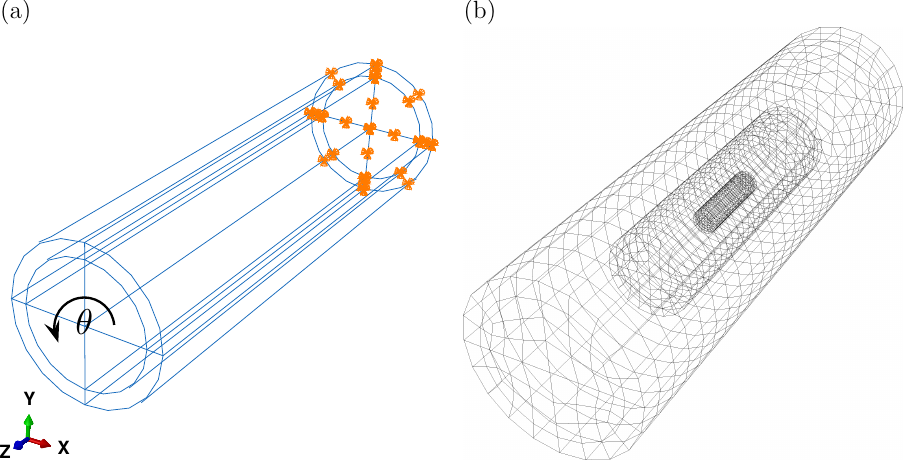} 
	\caption{(a) Boundary-value problem for  torsion of an fcc
          single crystal wire subjected to a prescribed  twist 
	angle $\theta$. In the calculations, the wire is taken to have
         a [001] orientation. (b) Example finite element meshes
  for wire diameters of $D=150$nm, 500nm, and 1000nm, nested into each other.} 
\label{fig:tor_FEM}
\end{figure}

Calculations were carried out on cylindrical specimens characterized by diameters $D$ ranging between $0.15\mu$m and $10\mu$m and a length to diameter ratio $L/D=3.5$.
Torsion boundary conditions were applied via an applied twist angle,
with $\nvec e_z$ being 
a unit vector along the axis of twist, \cref{fig:tor_FEM}a. The top
surface (at $z=L/2$) was subjected to twist angle $\theta$ while
all nodes of the bottom surface (at $z=-L/2$) were fixed,
\cref{fig:tor_FEM}a. A linear approximation was employed where
$\sin\theta \approx \theta$, suitable for small strain problems. The
specimens were loaded with a constant torsion rate of $\dot{\varphi}
\approx 5.6^{\circ}$ ns$^{-1}$.  

A relatively coarse mesh was used for all specimens, as illustrated in \cref{fig:tor_FEM}b. An analysis of mesh refinement and element type was carried out; see \cref{sec:discretization}. On that basis, most calculations were performed using meshes consisting of 1664 20 node brick elements with quadratic displacement shape functions and 27 point Gaussian integration.

\begin{figure}[h!]
	\centering
    \includegraphics[width=0.85\textwidth]{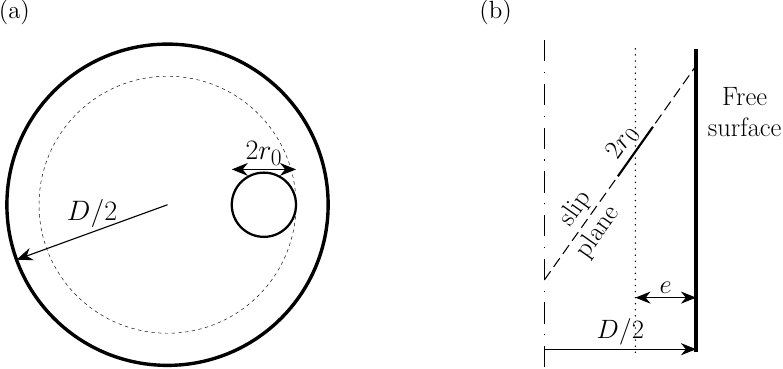} 
	\caption{Illustration of the exclusion distance $e$.  (a)
           Sketch of  a dislocation loop in a transverse plane of the wire; 
	(b)   Sketch of a dislocation loop on a slip plane inclined
           relative to the wire axis.}
\label{fig:sketch}
\end{figure}

The centers of dislocation sources were randomly dispersed along the
axial and hoop directions, but they were located radially at $D/2-e$
from the surface, where $e$ denotes an exclusion distance,
\cref{fig:sketch}. In the calculations here, 
$e=2r_0$ with $r_0$ being the initial loop radius. The number of sources was
adjusted so as to achieve a desired initial dislocation density
$\rho_0$. 
At nucleation, the radius of initial loops is scaled with the specimen size as $r_0=D/10$. Each loop is discretized using line element lengths $\xi = |\boldsymbol{\xi}_a|$ = 2 $\pi r_0/M(r_0)$. Also, no monopole splitting was used in any of the calculations.

A time step of $\Delta t = 0.3$ ps was chosen based on numerical
stability considerations \citep{ARIZA21}, derived from the smallest
dislocation loop ($r_0=15$ nm for a wire diameter of $D=150$ nm).

The material parameters were chosen to be representative of copper
but with isotropic elastic properties,
shear modulus $G = 48$GPa, and Poisson's ratio $\nu = 0.34$. The Burgers
vector length is taken to be $b = 0.2556$ nm, and lattice parameter is
$a =0.3614$ nm. Only dislocation glide is considered in the calculations and
the relation between glide component $f_{\rm glide}$ of ${\bf f}_a$
in Eq.~(\ref{HzEtbK}) and dislocation glide
velocity $v_{\rm glide}$ is taken to be independent of dislocation
character and given by
\begin{equation}
v_{\rm glide}=\frac{1}{B} f_{\rm glide}
\label{vglide}
\end{equation}
with the glide mobility constant taken to be $B = 10^{-5}$ Pa~s$^{-1}$. 

Results are presented  in terms of the normalized torque, $T/D^3$,
versus plastic strain on the surface, $\gamma^p_{\rm surf}$. The net
torque at any given time $t$ is calculated using: 
\begin{equation}
\nvec T = \sum_{\rm n=1}^{N_{\rm surfz+}} {\nvec x }_{\rm n} \times \hat{\nvec f}_{\rm n}
	+ \left[ \sum_{\rm n=1}^{N_{\rm surfz+}} {\nvec x }_{\rm n} \times \tilde{\nvec f}_{\rm n} 
	- \sum_{\rm n=1}^{N_{\rm surfz-}} {\nvec x }_{\rm n} \times \tilde{\nvec f}_{\rm n} \right]
\label{eq:s_consistenttorque}
\end{equation}
where ${\nvec x }_{\rm n}$ denotes the position of surface node $\rm n$ relative to the axis of torque, $N_{\rm surfz+}$ denotes the nodes corresponding to the positive $z$ face, and $N_{\rm surfz-}$ corresponds to the negative $z$ face. The first term in \cref{eq:s_consistenttorque} represents the net torque computed from the finite element solution, which is opposite and equal in magnitude to the reaction torque at the bottom surface. The second, bracketed term in \cref{eq:s_consistenttorque} represents a correction torque arising from the symmetry breaking due to random dislocation locations. 

The incremental plastic twist per unit length, $\Delta\theta^p$, due to dislocation motion, is obtained from 
\begin{equation}
\Delta\theta^p = \frac{\Delta\varphi}{L} = \sum_a^M \frac{1}{I_pL}-{(b_z^a \nu_x^a+b_x^a \nu_z^a)\bar{y}^a+ (b_z^a \nu_y^a+b_y^a \nu_z^a)\bar{x}^a}\Delta A^a_{swept}
\end{equation}
which is adapted from a similar equation used by \citet{RYU20}. Here,
$\bar{x}^a$ and $\bar{y}^a$ denote the centroid locations of the swept
area of the $a$th monopole element of line, $I_p= \pi R^4/2$ is the
polar moment of inertia, and $\nu_i$ are the components of the slip
plane normal. The surface shear strain, $\gamma^p_{\rm surf}$, is defined as
\begin{equation}
\gamma^p_{\rm surf} = \Delta\theta^p \left ( \frac{D}{2} \right )
	\label{eq:gps}
\end{equation}

\subsection{Numerical aspects}
\label{sec:discretization}

\begin{figure}[H]
    \centering
     \includegraphics[width=\textwidth]{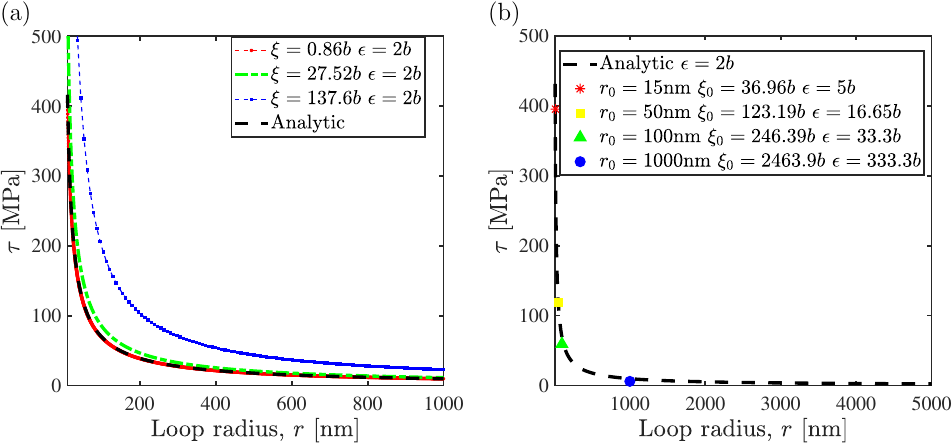} 
\caption{Plots of resolved shear stress, $\tau$, for a general dislocation loop, defined
  by \cref{eq:tau}  
        versus loop radius $r$.
	(a)  Comparison of the  monopole discretization results with
        the analytical solution of \citet{HirthBook82} for current
        line element values 
        $\xi=0.86$b, $\xi=27.52$b and $\xi=137.6$b. The value of the
        regularization parameter $\epsilon$ is fixed at $2$b.
	(b) The effect of the choice of initial values of  line
        element $\xi_0$ and  choice of the value of the regularization
        parameter $\epsilon$ for 
        dislocation loops with various values of initial radius  $r_0$.}
\label{fig:tau_gen_opt} 
\end{figure}

First, the optimal size of the element of dislocation line $\xi$,
needed to accurately reproduce the non-singular solution for
(isolated) general dislocation loops \citep{CAI06, HirthBook82} is
explored. \cref{fig:tau_gen_opt}a shows the shear stress, defined
based on the average Peach-Koehler force
\begin{equation}
    \tau = \frac{1}{M} \sum_a^M|\mathbf{f}_a^{\epsilon}|/b
    \label{eq:tau}
\end{equation}  
versus the loop radius for various values of $\xi$ using a core
regularization parameter $\epsilon=2b$. A value of $\xi=0.86b$ is
needed to recover the analytical solution of the general loop. For
larger loops, this value of $\xi$ leads to a  large
number of monopoles being required.
In order to reduce the computational time,
a scaling relation between $\xi$ and $\epsilon$ is used. This enables
calculations with larger values of $\xi$ to be used while maintaining
accuracy  as illustrated in \cref{fig:tau_gen_opt}b. 


In the calculations here, the values of $\xi$ and $\epsilon$ are set for $r_0=15$ nm and then proportionally scaled  with the dislocation loop radius. This maintains accuracy for the value of the resolved shear stress of a general loop of radius $r_0$ (\cref{fig:tau_gen_opt}b). As a result, only 10 monopoles are needed for both, a $r_0=15$ nm dislocation loop nucleated at wire diameters of $D=150$ nm and a $r_0=1000$ nm loop nucleated at $D=10000$ nm. This reduces the computational time needed for larger specimens by at least an order of magnitude.

\begin{figure}[h]
   \centering
    \includegraphics[width=\textwidth]{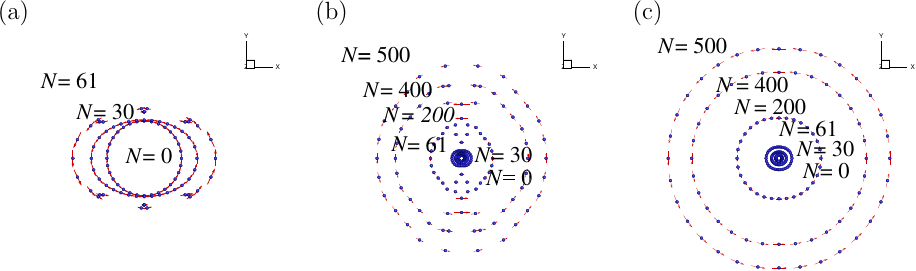} 
	\caption{Monopole configurations at various time increments
          $N$ for an initially circular 
 loop of radius $150 b$ subject to a remote shear stress using 
	(a) $\alpha = 1$, (b) $\alpha= 1.25$, and
	(c) $\alpha= 1.5$ in \cref{eq:nuc}.} 
\label{fig:nuc}
\end{figure}

Next, presume that loop nucleation occurs when the externally applied
stress exceeds the equilibrium shear stress of a loop with radius
$r_0$, as inferred from \cref{fig:tau_gen_opt}b. This would correspond
to using $\alpha = 1$ in \cref{eq:nuc}.  Calculations showed that this
condition can lead to high curvature-driven shape instabilities, as
illustrated in \cref{fig:nuc}a, causing the loop to partially collapse
and resulting in numerical convergence issues. Stable growth of
dislocation loops was found to occur using $\alpha \geq 1.5$. Some
examples are shown in \cref{fig:nuc}b and \cref{fig:nuc}c. In all subsequent
calculations  the value $\alpha = 1.5$ has been used.

\begin{figure}[h]
   \centering
    \includegraphics[width=\textwidth]{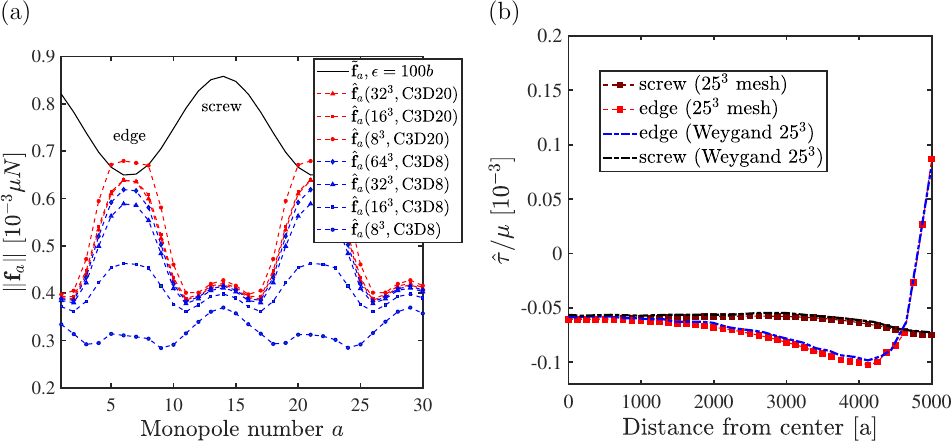} 
	   \caption{(a) Glide component of the Peach-Koehler force
             around a dislocation loop versus the number of monopoles
             used to represent the loop for
             various finite element mesh discretizations. The number
             in parenthesis is the number of elements per side of the
             cube and C3D20 and C3D8 denote 20 node and 8 node finite
             element shape functions, respectively.
	   (b) The image shear stress, $\hat{\tau}$, in the glide plane
             along a path that extends from the center of the
             dislocation  loop
             through the purely edge-oriented and screw-oriented parts
             of the dislocation to the free surface of the cube
             compared with the results of \citet{Weygand02} with a
             finite element discretization of $25$ elements per side. } 
           \label{fig:mono_conv_hat}
\end{figure}

The effect of free surfaces is now analyzed by comparing with a result
of  \citet{Weygand02} for a circular loop with radius
$r=4000a$, and Burgers vector $\mathbf{b}=a[100]/\sqrt{2}$, placed at
the center of a cube with side length $l=10000a$ that lies in the
$x-z$ plane. All surfaces of the cube are free of tractions. The loop
was discretized  using $\xi=1185b$ and $\epsilon=100b$. 

In these calculations, two types of spatial discretization were used:
20 node brick elements with quadratic displacement shape functions and
27 point Gaussian integration, as well as 8 node brick elements with
linear displacement shape functions and 8 point Gaussian integration. 

\cref{fig:mono_conv_hat}a shows the norm of the infinite body
Peach--Koehler force, $\tilde{\nvec f}_a$, and the norm of the finite
element computed image force,  $\hat{\nvec f}_a$, per monopole as 
the length of the loop is spanned starting at the screw segment. Results
are shown for the two types of finite elements and various mesh
densities denoted by the cube of the number of elements per side
of the cube. For reference, the $8$ elements per side and $64$
elements per side  correspond to
element sizes of $883b$ and $110b$, respectively. Using quadratic
elements (denoted by C3D20 in \cref{fig:mono_conv_hat}a), convergence is
attained for $16$ elements per side  (an element size of $441b \sim
100$nm), which is approximately
half the size of the line elements used to discretize the
loop. On the other hand, no convergence was obtained using the linear
elements (denoted by C3D8 in \cref{fig:mono_conv_hat}a ) up to $64$
elements per side.

It was verified that the hat component of the resolved shear stress,
$\hat{\tau}=-n_i\hat{\sigma}_{ij}b_j/b$, matches the numerical
solution obtained by \citet{Weygand02} using a line-based method, as
shown in Fig.~\ref{fig:mono_conv_hat}b.  
 
\section{Results}

\begin{figure}[h!]
     \centering
      \includegraphics[width=\textwidth]{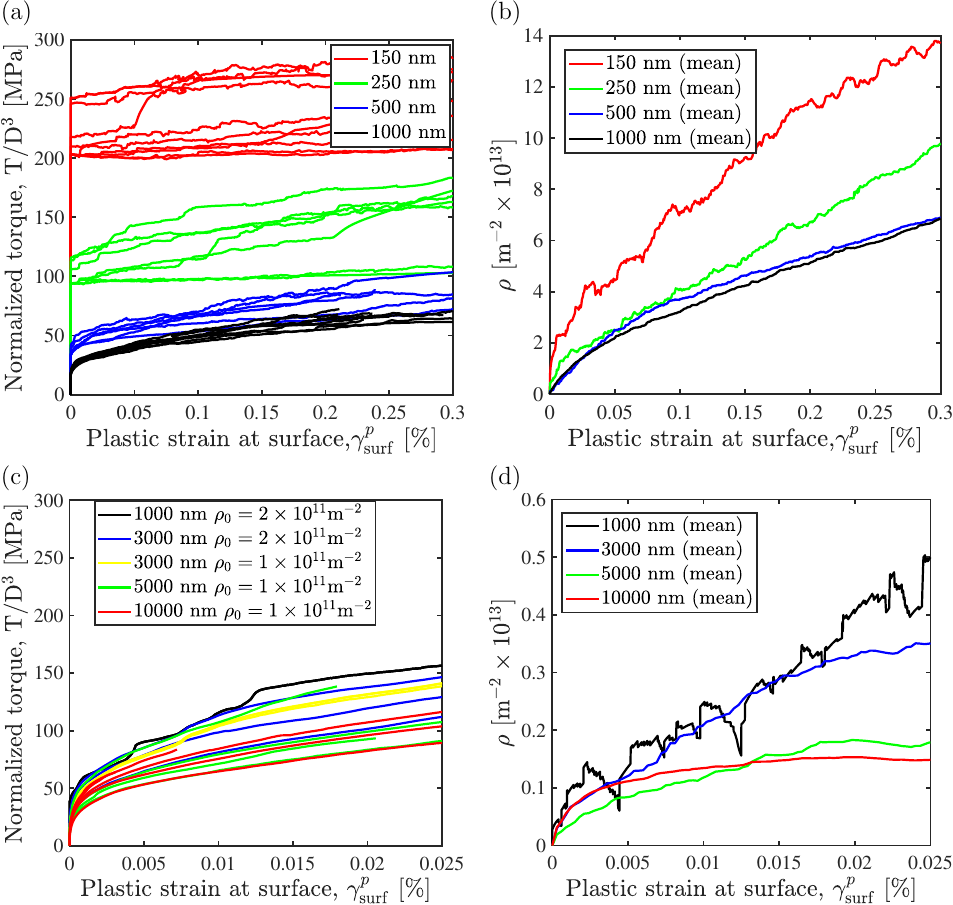} 
	   \caption{ (a)
             Normalized torque, $T/D^3$, versus surface plastic strain,
             $\gamma^p_{\rm surf}$, and  
             (b) dislocation density versus $\gamma^p_{\rm surf}$, for
             an initial dislocation density $\rho_0 = 6 \times 10^{13} \,
             \text{m}^{-2}$ and for wire diameters $D = 150, 250, 500,
             1000 $ nm. (c)  Normalized torque, $T/D^3$, versus surface plastic strain,
             $\gamma^p_{\rm surf}$, and  
             (b) dislocation density versus $\gamma^p_{\rm surf}$, for
             an initial dislocation density $\rho_0 = 0.01-0.02 \times 10^{13} \,
             \text{m}^{-2}$ and for wire diameters $D = 1000, 3000,
             5000, 10000 $ nm.} 
           \label{fig:torque_plast_strain}
\end{figure}

The normalized torque, $T/D^3$, where $T = |\nvec T \cdot \nvec e_z|$
is the torque calculated using \cref{eq:s_consistenttorque} and $D$ is
the wire diameter, is plotted in \cref{fig:torque_plast_strain} versus
the average surface plastic strain, $\gamma^p_{\rm surf}$, given by
\cref{eq:gps}. Calculations were carried out for 5 to 10 realizations
with the same wire diameter and initial  density of dislocation
sources.  In all calculations, the fcc single crystal wires are taken
to have a [001] orientation.
Two values of  initial dislocation source density were considered:
(i) an initial dislocation source density of $\rho_0 = 6 \times
10^{13}$ m$^{-2}$ for wire diameters in the range $D = 150$--$1000$
nm; and (ii)  a smaller value of  initial dislocation source
density, $\rho_0 = 1 \times 10^{11}$ m$^{-2}$, for wire diameters in
the range $D = 1000$--$10000$ nm. 

\cref{fig:torque_plast_strain}a shows the normalized torque versus
average plastic strain at the surface for wires with diameters in the
range between $D=150$ nm and $D=1000$ nm. For each wire diameter the
response is plotted for all realizations analyzed. The normalized
torque at $\gamma^p_{\rm surf}=0.2\%$ increases from about $60$ MPa for
$D=1000$ nm to $250$ MPa for $D=150$ nm with the variation in response
with realization being greater for smaller values of wire
diameter. This variation with wire diameter is similar to that found
by \cite{RYU20}. 

The evolution of the mean dislocation density $\rho$ for wire
diameters ranging from $D=150$ nm to $D=1000$ nm is shown in
Fig.~\ref{fig:torque_plast_strain}b. The formation of geometrically
necessary dislocations (GNDs) leads to a higher mean dislocation
density with increasing values of $\gamma^p_{\rm surf}$ and therefore
with torque $T$. The dislocation density increases more rapidly with
$\gamma^p_{\rm surf}$ for $D=150$ nm than for the wires with larger
diameters.  In the calculations here, the linear proportional relation
between dislocation density and the surface plastic strain, seen by
\citet{Senger11, RYU20} is not obtained.  

Figs.~\ref{fig:torque_plast_strain}c and
\ref{fig:torque_plast_strain}d, respectively, show the normalized
torque, $T/D^3$,   versus average surface plastic strain
$\gamma^p_{\rm   surf}$ response and the evolution of dislocation
density for wires 
with diameters in the range $D=1000$ nm to $D=10000$ nm. For  larger
diameter wires the curves of $T/D^3$ versus $\gamma^p_{\rm surf}$ show
less variation with realization and are smoother, i.e. there are fewer
strain bursts,  than the corresponding curves for smaller diameter
wires in Fig.~\ref{fig:torque_plast_strain}a. For these larger
diameter wires, there is still some size effect but it is much less
pronounced than for the smaller diameter wires. 

\begin{figure}[h]
   \centering
    \includegraphics[width=0.95\textwidth]{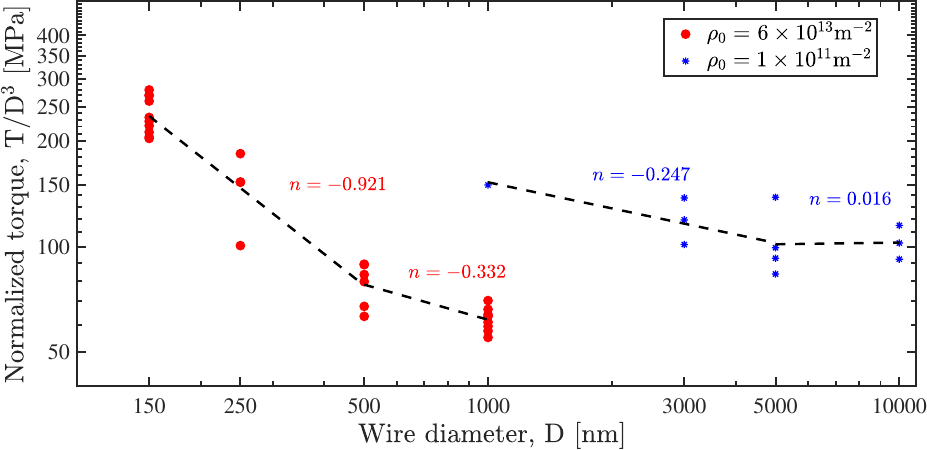} 
	   \caption{Log-log plots of  normalized torque, $T/D^3$,
             versus wire diameter $D$ for  single
             crystal fcc wires subject to a prescribed twist. Results are shown for 
             two values of initial dislocation
   density, $\rho_0$. For $\rho_0 = 6 \times 10^{13} \,
   \text{m}^{-2}$ the values of $T/D^3$ are at a  
             surface plastic strain of $\gamma^p_{\rm
               surf}=0.2\% $ with
             wire diameters of $D = 150, 250, 500, 1000 $ nm. For 
             $\rho_0 = 1\times 10^{11} \,
             \text{m}^{-2}$ the values of $T/D^3$ are at a  
             surface plastic strain of $\gamma^p_{\rm   surf}=0.02\%$
             with wire diameters of  $D = 1000, 3000,  
             5000, 10000 $ nm.} 
           \label{fig:torque_size_effect}
\end{figure}

The scaling of the normalized torque with size, i.e. 
with the wire diameter $D$, is shown in Fig.~\ref{fig:torque_size_effect}
which is a log-log plot of the average value over all realizations 
calculated of $T/D^3$ at a specified value of $\gamma^p_{\rm surf}$ versus the
wire diameter $D$.  One curve is for wires with diameters ranging 
from $D=150$ nm to $D=1000$ nm, with an initial dislocation 
source density of $\rho_0 = 6 \times 10^{13} \, \text{m}^{-2}$ and 
a torque evaluated at $\gamma^p_{\rm surf}=0.2\%$. The other curve is 
for  wires with  diameters ranging from $D=1000$ nm to $D=10000$ nm,
with an initial dislocation source density of  
$\rho_0 = 1 \times 10^{11} \, \text{m}^{-2}$
and with the torque evaluated at $\gamma^p_{\rm surf}=0.02\%$.

For both the smaller wires and the larger wires, the size dependence
can be characterized by a bilinear fit. For wire diameters in the
range from $D=150$ nm to $D=1000$ nm, for those with $150$ nm $\le D
\le $ $500$nm  the fit exponent is $n = -0.912$  and for wire
diameters in the range from $D=500$ nm to $D=1000$ nm  the fit
exponent is $n = -0.332$. While for the wires with the small values of
initial dislocation source density, which have diameters ranging from
$1000$ nm to $10000$ nm, the two exponents fit are $n = -0.247$ and
$n=0.016$, indicating that for a sufficiently large wire diameter, the
response is predicted to be nearly size independent. Thus, our
results indicate that the exponent characterizing the size effect in
torsion of single crystal wires is dislocation source density
dependent as well as size dependent. The values of the  exponents
characterizing the size dependence in previous discrete dislocation
plasticity analyses of wire torsion were $n = -0.96$ by
\cite{Senger11} and $n = -0.417$ by \cite{RYU20}. The largest wire
diameter considered by by
\cite{Senger11}  was 1$\mu$m and the largest wire diameter considered
by \cite{RYU20} was  2$\mu$m. 

\begin{figure}[h]
    \centering
    \includegraphics[width=0.95\textwidth]{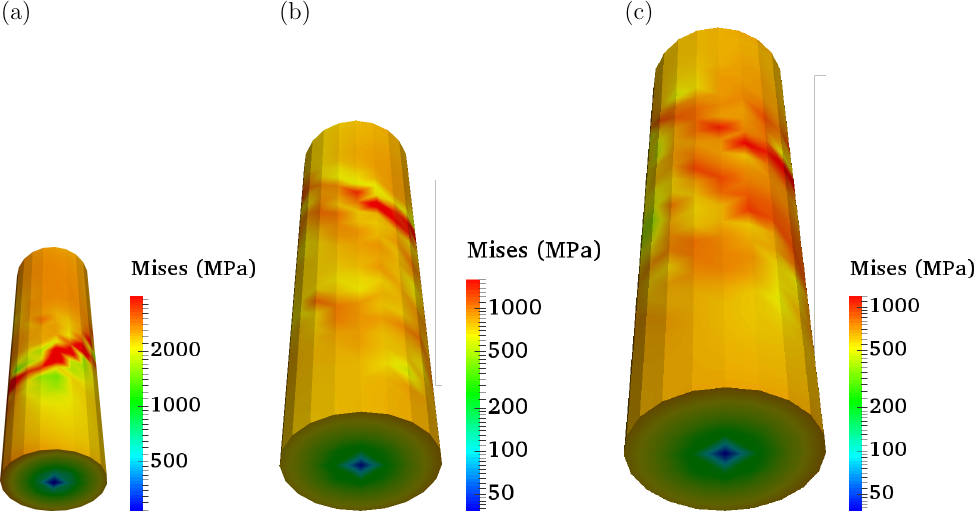} 
\caption{Contours of Mises effective stress, Eq.~(\ref{Mises}),  illustrating increasing
  localization in a band with decreasing wire diameter $D$  at
  $\gamma_{\rm surf}^p= 0.21 \%$ for  $\rho_0 = 6 \times 10^{13} \,
  \text{m}^{-2}$ and for  various wire diameters. (a) $D = 150$ 
  nm, (b) $D = 500$ nm and (c)  $D = 1000$ nm.} 
\label{fig:torque_size_contour}
\end{figure}

To gain insight into the bilinear size dependence, contours of the von
Mises effective stress, defined by 
\begin{equation}
\sigma_e=\sqrt{{\boldsymbol{\sigma}}^\prime :
    {\boldsymbol{\sigma}}^\prime} \quad , \quad
 {\boldsymbol{\sigma}}^\prime= {\boldsymbol{\sigma}}-\frac{1}{3} {\rm tr}(
 {\boldsymbol{\sigma}} ) {\bf I},
\label{Mises}
\end{equation}
are shown in \cref{fig:torque_size_contour} at $\gamma^p_{\rm surf} = 0.21\%$
for wires with three values of wire diameter, $D=150$ nm in
Fig.~\ref{fig:torque_size_contour}a, $D=500$ nm in
Fig.~\ref{fig:torque_size_contour}b and $D=1000$ nm in
Fig.~\ref{fig:torque_size_contour}c.
Recall that the meshes used are relatively coarse so that the stresses are actually
more concentrated than shown in the figure.
The main effect in \cref{fig:torque_size_contour}  is that the high
values of Mises effective stress are in a single band for $D=150$ nm
and that the distribution of Mises effective stress becomes more
uniformly distributed  for the larger values of $D$. The change in
slope in \cref{fig:torque_size_effect} is associated with the change
from a nucleation-controlled, localized distribution of plasticity to
one that is more uniformly distributed. Also, for all three values of
wire diameter, there is a radial gradient of the Mises effective
stress, which varies from near zero in the center of the wire to its
maximum value at the surface.  

\begin{figure}[h!]
    \centering
     \includegraphics[width=0.80\textwidth]{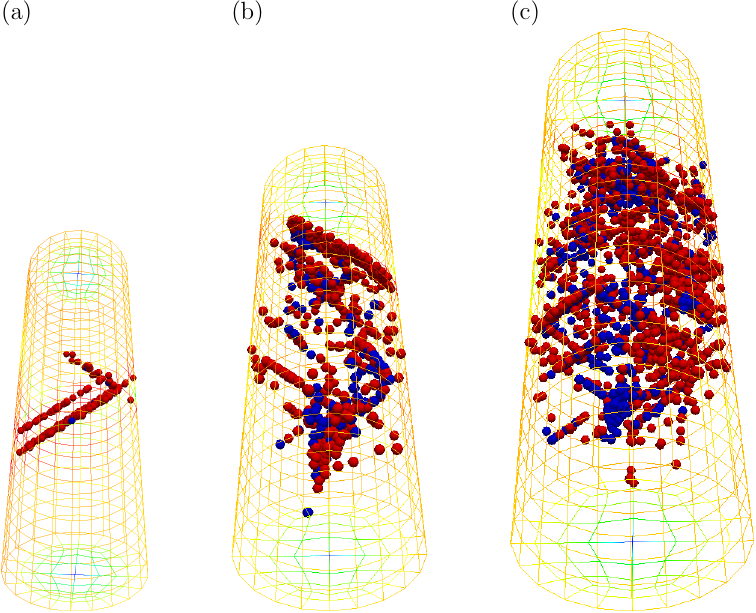} 
\caption{Monopole locations showing the dislocation network with
  $\rho_0 = 6 \times 10^{13} \,   \text{m}^{-2}$  at
  $\gamma_{\text{\rm surf}}^p = 0.21\%$  and with  wire diameters: (a) $D =
  150$ nm, (b) $D = 500$ nm, and (c) $D = 1000$ nm. 
The dark blue colored monopoles denote dislocations that
have velocity magnitudes less than $20$ m/s and that are
essentially immobile. 
}
\label{fig:torque_size_dislo}
\end{figure} 
\begin{figure}[h]
    \centering
     \includegraphics[width=\textwidth]{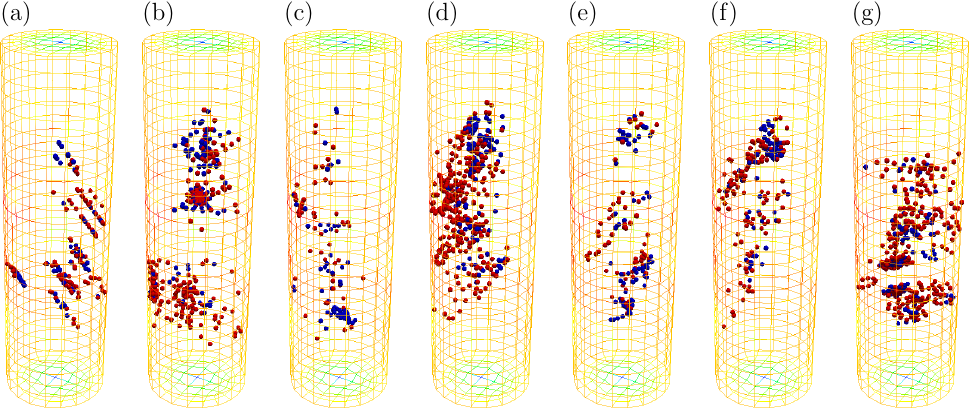} 
\caption{Monopole locations showing the dislocation  configurations on
  active slip systems at 
  $\gamma_{\rm surf}^p= 0.21\%$ with $D=1000$nm and $\rho_0 = 6
  \times 10^{13} \,   \text{m}^{-2}$. (a) 
  $\{111\}\langle\bar 101\rangle$, (b) $\{1\bar
  11\}\langle011\rangle$, (c) $\{1\bar 11\}\langle\bar 101\rangle$,
  (d) $\{\bar 1\bar 11\}\langle1\bar 10\rangle$, (e) $\{\bar 1\bar
  11\}\langle011\rangle$, (f) $\{\bar 1\bar 11\}\langle101\rangle$,
  (g) $\{\bar 111\}\langle110\rangle$.  The dark colored monopoles
  denote dislocations that 
have velocity magnitudes less than  20 m/s and that are
essentially immobile. }
\label{fig:torque_planes}
\end{figure} 

\cref{fig:torque_size_dislo} shows the dislocation structure for the
three cases in Fig.~\ref{fig:torque_size_contour}. With $D=150$nm
(Fig. \ref{fig:torque_size_dislo}a), only a few dislocation sources on
a few glide planes are activated initially. With increasing twist,
plastic deformation predominantly involves source nucleation, and then
subsequently dislocation exhaustion occurs due to dislocations exiting
the free surface \citep{Greer05,BenzergaIJP}. This can lead to an
abrupt increase in the value of normalized torque, $T/D^3$, due to the
inability to nucleate more dislocation loops, thus leading to a higher
stress level.  The dark blue colored monopoles denote dislocations that
have velocity magnitudes less than 20 m/s and that are
essentially immobile.  Immobile dislocations in pile-ups occur  at the
center of the wire. The mobile dislocations can attain very high
velocity magnitudes due to the linear mobility law, Eq.~(\ref{vglide}),
and the high stresses that occur in the wires.

With $D=1000$ nm, \cref{fig:torque_size_dislo}c, more dislocation
planes and sources are activated, leading to complex interactions
between the mobile and immobile dislocations. These results are
consistent with those of \citet{Senger11} on wires having a
$\langle100\rangle$ orientation when cross-slip was not activated,
showing no long dislocation segments along the torsion axis and that
dislocation pile-ups occur perpendicular to it.

Further details about the dislocation configuration are given in
\cref{fig:torque_planes} for the $D=1000$nm wire of
\cref{fig:torque_size_dislo}c using the same dislocation velocity
scale. Here, dislocation monopoles on individual (active) slip systems
are shown at $\gamma_{\rm surf}^p$ = 0.21 $\%$. The largest
dislocation density occurs on the $\{1\bar 11\} \langle011\rangle$,
${\bar 1\bar 11}\langle1\bar 10\rangle$, and  ${\bar
  111}\langle110\rangle$ slip planes. Due to the  high dislocation
activity near the surface of the wire,  the  fast-moving dislocations
(red) lead to dislocations exiting the wire, inducing dislocation
starvation. On other slip planes, such as ${1\bar 11}\langle\bar
101\rangle$, ${\bar 1\bar 11}\langle011\rangle$, and  ${\bar 1\bar
  11}\langle101\rangle$, there are trapped dislocations near the
center of the wire, leading to dislocations being immobile there.

\section{Concluding remarks}

A three-dimensional, small strain discrete dislocation plasticity
framework for solving quasi-static finite body  boundary value
problems  has been 
presented that  combines the monopole representation 
of dislocations with  image fields that enforce the boundary
conditions.  In this framework,
both the overall plastic flow and the evolving dislocation structure
are outcomes of the boundary value problem solution. 
Torsional
loading calculations were carried out for wires with diameters
varying over three orders of magnitude and 
for two values of the initial dislocation density. The specimen size
dependence and the dislocation structures that 
develop were investigated for  single crystal wires
subject to torsional loading.  The calculations indicate that:

\begin{itemize}

\item There is a  strong size-dependence of the torque, normalized by the cube
  of the wire diameter, 
  with the scaling exponent of the size-dependence 
  being dependent on specimen size with a less strong dependence  on
  initial dislocation   density. 

\item There is a transition with increasing wire diameter  from
  nucleation controlled to
  interaction controlled dislocation plasticity with a
  strong size dependence in both these regimes. The
  interaction regime is dominated by the 
  emergence of geometrically necessary dislocations and dislocation
  pile-ups perpendicular to the axis of twist.  For wires with a
  sufficiently large wire diameter,  plastic
  deformation in the wire is much less heterogeneous and the
  dependence on size is greatly 
  reduced. 

\item The capabilities of the discrete dislocation plasticity
  framework developed here can be enhanced by  incorporating
  topological dislocation reactions leading to junction 
  formation.  

\item The  dislocation dynamics code used in the
  calculations was developed to explore the capabilities of the
  monopole representation. Computational efficiency can be much
  improved by parallelization of the code and by using a fast multipole method
  for the long range dislocation interaction calculations.
 
\end{itemize}

\section*{Acknowledgments}
AAB acknowledges support from NSF under grant CMMI-1950027. AC and AAB
thank Vincent Chiaruttini and Jean-Didier Garaud from ONERA for
assistance with the plugin interface of Z-set. 

\begin{appendices}
\renewcommand{\theequation}{\thesection-\arabic{equation}}
\renewcommand\thefigure{\thesection-\arabic{figure}}
\setcounter{figure}{0}
\crefalias{section}{appendix}

\setcounter{equation}{0} 
\section{The elastic field of dislocation monopoles} \label{app:monelas}

For completeness, we collect the monopole velocity and stress fields, Eq.~(\ref{e:sigma_tilde}), resulting from classical and regularized linear elasticity. 

\subsection{Green's function}
The displacement field ${\nvec u}(\nvec x)$ of an infinite solid loaded by body forces ${\nvec f}(\nvec x)$ is a solution of Navier's equation
\begin{equation}
    \frac{\partial}{\partial x_j} 
    \Big( c_{ijkl} \frac{\partial u_k}{\partial x_l}(\nvec x) \Big)
    +
    f_i(\nvec x)
    =
    0 ,
\end{equation}
where $c_{ijkl} = c_{klij} = c_{jikl} = c_{ijlk}$ are the elastic moduli of the solid. The Green's function $G_{ij}(\nvec x)$ is the $i$th component of the elastic field corresponding to a point load at the origin acting in the $j$th direction, i.~e.,
\begin{equation}
    \frac{\partial}{\partial x_j} 
    \Big( c_{ijkl} \frac{\partial G_{km}}{\partial x_l}(\nvec x) \Big)
    +
    \delta_{im} \, \delta(\nvec x)
    =
    0 ,
\end{equation}
where $\delta(\nvec x)$ is Dirac's delta. 
For an isotropic elastic solid, the Green's function follows explicitly as
\begin{equation}
    G_{ij}(\nvec x)
    =
    \frac{1}{8\pi\mu}
    \left(
        \delta_{ij} \nabla^2 r
        -
        \frac{\lambda+\mu}{\lambda+2\mu} 
        \frac{\partial^2 r}{\partial x_i \partial x_j}
    \right) ,
\end{equation}
where $r := |\nvec x|$, $\nabla^2$ is the Laplacian and $\lambda$ and $\mu$ are the Lam\'e constants. 

\subsection{Displacement and stress fields}
Consider a dislocation line in the form of a closed line $C$. The Burgers vector $b_i(\nvec x)$ can change value at branching points of $C$, but it must be conserved according to Frank's rule. Then, Mura's formula (\citet{Mura:1963}; cf.~\citet[eq.~(4-12)]{HirthBook82}) gives the displacement gradient as 
\begin{equation} \label{lBN90s}
    \frac{\partial u_m}{\partial x_s}(\nvec x)
    = 
    \oint_C 
        c_{ijkl} 
        \frac{\partial G_{mk}}{\partial x'_l}(\nvec x' - \nvec x) 
        {\rm e}_{jsn} b_i(\nvec x')
    dx'_n ,
\end{equation}
whence the corresponding stress field follows from Hooke's law as
\begin{equation} \label{7eVmpk}
    \sigma_{ij}(\nvec x)
    =
    c_{ijkl} \frac{1}{2}
    \left(
        \frac{\partial u_k}{\partial x_l}(\nvec x)
        +
        \frac{\partial u_l}{\partial x_k}(\nvec x)
    \right) ,
\end{equation}
and the displacement field follows by Burgers' formula (
cf.~\citet[eq.~(4-6)]{HirthBook82}) as
\begin{equation} \label{a7CF3T}
    u_m(\nvec x)
    =
    -
    \int_S
        c_{ijkl} 
        \frac{\partial G_{mk}}{\partial x'_l}(\nvec x' - \nvec x)
        b_i(\nvec x')
    \, dS_j(\nvec x') ,
\end{equation}
where $dS_j(\nvec x)$ denotes the element of oriented area at $\nvec
x$. We note that the evaluation of Eq.~(\ref{a7CF3T}) requires
explicit knowledge of the slip surface $S$ and cannot be reduced to a
line integral. By contrast, the velocity field due to a moving
dislocation line does reduce properly to a line integral. Thus, taking
rates in Eq.~(\ref{a7CF3T}) and applying Orowan's relation we obtain 
\begin{equation} \label{HIb7D1} 
    \dot{u}_m(\nvec x)
    =
    -
    \oint_C
        c_{ijkl} 
        \frac{\partial G_{mk}}{\partial x'_l}(\nvec x' - \nvec x)
        {\rm e}_{jpq}
        b_i(\nvec x') v_p(\nvec x')
    \, dx_q(\nvec x') ,
\end{equation}
where $v_p(\nvec x)$ denotes the velocity of the dislocation line $C$ at $\nvec x$. 

\subsection{Monopole approximation}

Approximating the dislocation distribution by a collection of monopoles and using the approximating property Eq.~(\ref{1QMIwo}), the classical integral expressions of the stress and velocity fields, Eq.~(\ref{lBN90s}) and Eq.~(\ref{HIb7D1}), become
\begin{equation} \label{hQaA0W}
    \sigma_{pq}(\nvec x)
    = 
    \sum_{a=1}^M 
        c_{pqms} c_{ijkl} 
        \frac{\partial G_{mk}}{\partial x'_l}(\nvec{x}_a - \nvec x) 
        {\rm e}_{jsn} b_{a i}
    \xi_{a n} ,
\end{equation}
and
\begin{equation} \label{IuizhN}
    \dot{u}_m(\nvec x)
    =
    -
    \sum_{a=1}^M 
        c_{ijkl} 
        \frac{\partial G_{mk}}{\partial x'_l}(\nvec{x}_a - \nvec x)
        {\rm e}_{jpq}
        b_{a i} v_{a p}
    \, \xi_{a q}  ,
\end{equation}
respectively,  where $v_{a p} := (\nvec{v}_a)_p$ is the velocity of monopole $a$. Thus, the discrete stress and velocity fields follow as the sum of monopole contributions, as surmised in Eq.~(\ref{e:sigma_tilde}) and expected from linearity. 

\subsection{Core regularization}

The stresses $\nten{\sigma}(\nvec{x})$ computed directly from linear elasticity as in Eq.~(\ref{hQaA0W}) diverge as $\nvec{x}$ approaches one of the monopoles and Eq.~(\ref{hQaA0W}) cannot be used directly to evaluate the self-forces of the monopoles in Eq.~(\ref{HzEtbK}). One way to overcome this difficulty is to regularize linear elasticity and replace it by strain-gradient elasticity of the Mindlin type \citep{Mindlin:1964}. A careful analysis \citep{Deffo19} shows that this regularization is equivalent to replacing the Green's function $G$ of conventional linear elasticity in Eq.~(\ref{hQaA0W}) and Eq.~(\ref{IuizhN}) by a regularized Green's function
\begin{equation} \label{cfl5gE}
    G^\epsilon_{ij}(\nvec{x})
    =
    \iint
        G_{ij}(\nvec{x}-\nvec{x}'-\nvec{x}'') 
        \varphi^\epsilon(\nvec{x}')
        \varphi^\epsilon(\nvec{x}'')
    \, dV(\nvec{x}') \, dV(\nvec{x}'') ,
\end{equation}
where $\epsilon$ is commensurate with the crystal lattice parameter and
\begin{equation}
    \varphi^\epsilon(\nvec{x})
    =
    \frac{1}{4\pi \epsilon^2 r}
    \,
    {\rm e}^{- r/\epsilon } ,
    \qquad
    r := |\nvec{x}| ,
\end{equation}
represents the core structure of the monopoles in Mindlin strain-gradient elasticity. For isotropic linear elasticity, Eq.~(\ref{cfl5gE}) evaluates explicitly to \citep{Deffo19} 
\begin{equation}
    G^\epsilon_{ij}(\nvec{x})
    =
    \frac{1}{8\pi\mu}
    \left(
        \nabla^2 r^\epsilon \delta_{ij}
        -
        \frac{\lambda+\mu}{\lambda+2\mu} 
        \frac{\partial^2 r^\epsilon}{\partial x_i \partial x_j}
    \right) ,
\end{equation}
with
\begin{equation}
    r^\epsilon
    :=
    (r^2 + 4 \epsilon^2 - \epsilon (r + 4 \epsilon){\rm e}^{-r/\epsilon})/r .
\end{equation}
The regularized Green's function $G^\epsilon_{ij}(\nvec{x})$ is finite
for $\epsilon > 0$. It can be readily verified that
$G^\epsilon_{ij}(\nvec{x})$ converges to $G_{ij}(\nvec{x})$ in a
distributional sense as $\epsilon\to 0$. In particular, the
regularized stresses and velocities converge pointwise to the
corresponding linear elastic fields in the same limit.   

\end{appendices}


\begin{thebibliography}{47}
\providecommand{\natexlab}[1]{#1}

\bibitem[{Ariza \protect\BIBand{} Ortiz(2021)}]{ARIZA21}
Ariza, M.P., Ortiz, M., 2021.
\newblock A semi-discrete line-free method of monopoles for dislocation
  dynamics.
\newblock Extreme Mechanics Letters 45, 101267.

\bibitem[{Arroyo \protect\BIBand{} Ortiz(2006)}]{ArroyoOrtiz:2006}
Arroyo, M., Ortiz, M., 2006.
\newblock Local maximum-entropy approximation schemes: a seamless bridge
  between finite elements and meshfree methods.
\newblock International Journal for Numerical Methods in Engineering 65,
  2167--2202.

\bibitem[{Arsenlis et~al.(2007)Arsenlis, Cai, Tang, Rhee, Oppelstrup, Hommes,
  Pierce, \protect\BIBand{} Bulatov}]{Arsenlis07}
Arsenlis, A., Cai, W., Tang, M., Rhee, M., Oppelstrup, T., Hommes, G., Pierce,
  T.G., Bulatov, V.V., 2007.
\newblock Enabling strain hardening simulations with dislocation dynamics.
\newblock Modell. Simul. Mater. Sci. Eng. 15, 553--595.

\bibitem[{Ashby(1970)}]{Ashby70}
Ashby, M.F., 1970.
\newblock The deformation of plastically non-homogeneous materials.
\newblock Philos. Mag. 21, 399--424.

\bibitem[{Benzerga(2008)}]{BenzergaIJP}
Benzerga, A.A., 2008.
\newblock An analysis of exhaustion hardening in micron-scale plasticity.
\newblock Int. J. Plast. 24, 1128--1157.

\bibitem[{Bertin et~al.(2015)Bertin, Upadhyay, Pradalier, \protect\BIBand{}
  Capolungo}]{Bertin15}
Bertin, N., Upadhyay, M.V., Pradalier, C., Capolungo, L., 2015.
\newblock {A FFT-based formulation for efficient mechanical fields computation
  in isotropic and anisotropic periodic discrete dislocation dynamics}.
\newblock Modell. Simul. Mater. Sci. Eng. 23, 065009.

\bibitem[{Bulatov et~al.(1998)Bulatov, Abraham, Kubin, Devincre,
  \protect\BIBand{} Yip}]{Bulatov98}
Bulatov, V.V., Abraham, F.F., Kubin, L., Devincre, B., Yip, S., 1998.
\newblock Connecting atomistic and mesoscale simulations of crystal plasticity.
\newblock Nature 391, 669--672.

\bibitem[{Bulatov \protect\BIBand{} Cai(2006)}]{BulatovCai:2006}
Bulatov, V.V., Cai, W., 2006.
\newblock Computer simulations of dislocations.
\newblock Oxford series on materials modelling. Oxford University Press,
  Oxford; New York.

\bibitem[{Cai et~al.(2006)Cai, Arsenlis, Weinberger, \protect\BIBand{}
  Bulatov}]{CAI06}
Cai, W., Arsenlis, A., Weinberger, C.R., Bulatov, V.V., 2006.
\newblock A non-singular continuum theory of dislocations.
\newblock Journal of the Mechanics and Physics of Solids 54, 561--587.

\bibitem[{Carrillo et~al.(2019)Carrillo, Craig, \protect\BIBand{}
  Patacchini}]{carrillo2019}
Carrillo, J.A., Craig, K., Patacchini, F.S., 2019.
\newblock A blob method for diffusion.
\newblock Calculus of Variations and Partial Differential Equations 58, 53.

\bibitem[{Carrillo et~al.(2017)Carrillo, Huang, Patacchini, \protect\BIBand{}
  Wolansky}]{carrillo2017}
Carrillo, J.A., Huang, Y., Patacchini, F.S., Wolansky, G., 2017.
\newblock Numerical study of a particle method for gradient flows.
\newblock Kinetic and Related Models 10, 613--641.

\bibitem[{Cleveringa et~al.(1999)Cleveringa, Van~der Giessen, \protect\BIBand{}
  Needleman}]{Cleveringa99}
Cleveringa, H.H.M., Van~der Giessen, E., Needleman, A., 1999.
\newblock {A Discrete Dislocation Analysis of Bending}.
\newblock Int. J. Plast. 15, 837--868.

\bibitem[{Cleveringa et~al.(2000)Cleveringa, Van~der Giessen, \protect\BIBand{}
  Needleman}]{Cleveringa00}
Cleveringa, H.H.M., Van~der Giessen, E., Needleman, A., 2000.
\newblock {A discrete dislocation analysis of mode I crack growth}.
\newblock J. Mech. Phys. Solids 48, 1133--1157.

\bibitem[{Comer(1979)}]{Comer79}
Comer, D., 1979.
\newblock {The ubiquitous B-tree}.
\newblock Computing Surveys 11, 121--137.

\bibitem[{Crone et~al.(2014)Crone, Chung, Leiter, Knap, Aubry, Hommes,
  \protect\BIBand{} Arsenlis}]{Crone14}
Crone, J.C., Chung, P.W., Leiter, K.W., Knap, J., Aubry, S., Hommes, G.,
  Arsenlis, A., 2014.
\newblock A multiply parallel implementation of finite element-based discrete
  dislocation dynamics for arbitrary geometries.
\newblock Modelling and Simulation in Materials Science and Engineering 22,
  035014.

\bibitem[{Deffo et~al.(2019)Deffo, Ariza, \protect\BIBand{} Ortiz}]{Deffo19}
Deffo, A., Ariza, M.P., Ortiz, M., 2019.
\newblock A line-free method of monopoles for 3d dislocation dynamics.
\newblock Journal of the Mechanics and Physics of Solids 122, 566 -- 589.

\bibitem[{Deng et~al.(2008)Deng, El-Azab, \protect\BIBand{} Larson}]{Deng08}
Deng, J., El-Azab, A., Larson, B., 2008.
\newblock On the elastic boundary value problem of dislocations in bounded
  crystals.
\newblock Philosophical Magazine 88, 3527--3548.

\bibitem[{El-Awady(2015)}]{Awady15}
El-Awady, J.A., 2015.
\newblock Unravelling the physics of size-dependent dislocation-mediated
  plasticity.
\newblock Nature Communications 6, 5926.

\bibitem[{Fedeli et~al.(2017)Fedeli, Pandolfi, \protect\BIBand{}
  Ortiz}]{Fedeli:2017}
Fedeli, L., Pandolfi, A., Ortiz, M., 2017.
\newblock Geometrically exact time-integration mesh-free schemes for
  advection-diffusion problems derived from optimal transportation theory and
  their connection with particle methods.
\newblock International Journal for Numerical Methods in Engineering 112,
  1175--1193.

\bibitem[{Fleck et~al.(1994)Fleck, Muller, Ashby, \protect\BIBand{}
  Hutchinson}]{Fleck94}
Fleck, N.A., Muller, G.M., Ashby, M.F., Hutchinson, J.W., 1994.
\newblock Strain gradient plasticity - theory and experiment.
\newblock Acta Metall. Mater. 42, 475--487.

\bibitem[{Ghoniem et~al.(2000)Ghoniem, Tong, \protect\BIBand{} Sun}]{Ghoniem00}
Ghoniem, N.M., Tong, S.H., Sun, L.Z., 2000.
\newblock Parametric dislocation dynamics: A thermodynamics-based approach to
  investigations of mesoscopic plastic deformation.
\newblock Phys. Rev. B 61, 913--927.

\bibitem[{Gravell \protect\BIBand{} Ryu(2020)}]{GRAVELL2020}
Gravell, J.D., Ryu, I., 2020.
\newblock Latent hardening/softening behavior in tension and torsion combined
  loadings of single crystal fcc micropillars.
\newblock Acta Materialia 190, 58--69.

\bibitem[{Greer et~al.(2005)Greer, Oliver, \protect\BIBand{} Nix}]{Greer05}
Greer, J.R., Oliver, W.C., Nix, W.D., 2005.
\newblock Size dependence of mechanical properties of gold at the micron scale
  in the absence of strain gradients.
\newblock Acta Mater. 53, 1821--1830.

\bibitem[{Hirth \protect\BIBand{} Lothe(1982)}]{HirthBook82}
Hirth, J.P., Lothe, J., 1982.
\newblock {Theory of Dislocations}.
\newblock Wiley, New York.

\bibitem[{Hutchinson(2000)}]{Hutchinson00}
Hutchinson, J.W., 2000.
\newblock Plasticity at the micron scale.
\newblock Int. J. Solids Struct. 37, 225--238.

\bibitem[{Joa et~al.(2023)Joa, Dupuy, R{\aa}back, Fivel, Perez,
  \protect\BIBand{} Amodeo}]{Joa23}
Joa, J.A.G., Dupuy, L., R{\aa}back, P., Fivel, M., Perez, M., Amodeo, J., 2023.
\newblock {El-Numodis: a new tool to model dislocation and surface
  interactions}.
\newblock Modell. Simul. Mater. Sci. Eng. 31, 055003.

\bibitem[{Kysar et~al.(2007)Kysar, Gan, Morse, Chen, \protect\BIBand{}
  Jones}]{Kysar07}
Kysar, J.W., Gan, Y.X., Morse, T.L., Chen, X., Jones, M.E., 2007.
\newblock High strain gradient plasticity associated with wedge indentation
  into face-centered cubic crystals: {G}eometrically necessary dislocation
  densities.
\newblock J. Mech. Phys. Solids 55, 1554--1573.

\bibitem[{Lazar(2017)}]{Lazar:2017}
Lazar, M., 2017.
\newblock Non-singular dislocation continuum theories: strain gradient
  elasticity vs. peierls–nabarro model.
\newblock Philosophical Magazine 97, 3246--3275.

\bibitem[{Leiter et~al.(2013)Leiter, Crone, \protect\BIBand{} Knap}]{LEITER13}
Leiter, K.W., Crone, J.C., Knap, J., 2013.
\newblock An algorithm for massively parallel dislocation dynamics simulations
  of small scale plasticity.
\newblock Journal of Computational Science 4, 401 -- 411.

\bibitem[{Liu et~al.(2005)Liu, Huang, Li, Hwang, \protect\BIBand{} Liu}]{Liu05}
Liu, B., Huang, Y., Li, M., Hwang, K.C., Liu, C., 2005.
\newblock A study of the void size effect based on the taylor dislocation
  model.
\newblock International Journal of Plasticity 21, 2107--2122.

\bibitem[{Mindlin(1964)}]{Mindlin:1964}
Mindlin, R.D., 1964.
\newblock Micro-structure in linear elasticity.
\newblock Arch. Rational Mech. Anal. 16, 51--78.

\bibitem[{Mura(1963)}]{Mura:1963}
Mura, T., 1963.
\newblock Continuous distribution of moving dislocations.
\newblock Phil. Mag. 8, 843--859.

\bibitem[{Mura(1982)}]{Mura82}
Mura, T., 1982.
\newblock {Micromechanics of Defects in Solids}.
\newblock {Martinus Nijhoff Publishers}.

\bibitem[{Nicola et~al.(2006)Nicola, Xiang, Vlassak, Van~der Giessen,
  \protect\BIBand{} Needleman}]{Nicola06}
Nicola, L., Xiang, Y., Vlassak, J.J., Van~der Giessen, E., Needleman, A., 2006.
\newblock Plastic deformation of freestanding thin films: Experiments and
  modeling.
\newblock J. Mech. Phys. Solids 54, 2089--2110.

\bibitem[{Nix \protect\BIBand{} Gao(1998)}]{Nix98}
Nix, W.D., Gao, H., 1998.
\newblock Indentation size effects in crystalline materials: A law for strain
  gradient plasticity.
\newblock J. Mech. Phys. Solids 46, 411--425.

\bibitem[{Nye(1953)}]{nye:1953}
Nye, J.F., 1953.
\newblock {Some geometrical relations in dislocated crystals}.
\newblock Acta Metallurgica 1, 153--162.

\bibitem[{Pandolfi et~al.(2023)Pandolfi, Stainier, \protect\BIBand{}
  Ortiz}]{Pandolfi:2023}
Pandolfi, A., Stainier, L., Ortiz, M., 2023.
\newblock An optimal-transport finite-particle method for mass diffusion.
\newblock Computer Methods in Applied Mechanics and Engineering 416, 116385.

\bibitem[{Ryu et~al.(2020)Ryu, Gravell, Cai, Nix, \protect\BIBand{}
  Gao}]{RYU20}
Ryu, I., Gravell, J., Cai, W., Nix, W.D., Gao, H., 2020.
\newblock Intrinsic size dependent plasticity in bcc micro-pillars under
  uniaxial tension and pure torsion.
\newblock Extreme Mechanics Letters 40, 100901.

\bibitem[{Senger et~al.(2011)Senger, Weygand, Kraft, \protect\BIBand{}
  Gumbsch}]{Senger11}
Senger, J., Weygand, D., Kraft, O., Gumbsch, P., 2011.
\newblock Dislocation microstructure evolution in cyclically twisted
  microsamples: a discrete dislocation dynamics simulation.
\newblock Modell. Simul. Mater. Sci. Eng. 19, 074004.

\bibitem[{St{\"{o}}lken \protect\BIBand{} Evans(1998)}]{Stolken98}
St{\"{o}}lken, J.S., Evans, A.G., 1998.
\newblock A microbend test method for measuring the plasticity length scale.
\newblock Acta Mater. 46, 5109--5115.

\bibitem[{Uchic et~al.(2009)Uchic, Shade, \protect\BIBand{} Dimiduk}]{Uchic09}
Uchic, M.D., Shade, P.A., Dimiduk, D.M., 2009.
\newblock Plasticity of micrometer-scale single crystals in compression.
\newblock Annu. Rev. Mater. Res 39, 361--386.

\bibitem[{{Van}~der Giessen \protect\BIBand{} Needleman(1995)}]{Giessen95}
{Van}~der Giessen, E., Needleman, A., 1995.
\newblock Discrete dislocation plasticity: a simple planar model.
\newblock Modell. Simul. Mater. Sci. Eng. 3, 689--735.

\bibitem[{Vattr\'e et~al.(2014)Vattr\'e, Devincre, Feyel, Gatti, Groh, Jamond,
  \protect\BIBand{} Roos}]{Vattre14}
Vattr\'e, A., Devincre, B., Feyel, F., Gatti, R., Groh, S., Jamond, O., Roos,
  A., 2014.
\newblock {Modelling crystal plasticity by 3D dislocation dynamics and the
  finite element method: The Discrete-Continuous Model revisited}.
\newblock J. Mech. Phys. Solids 63, 491--505.

\bibitem[{Weinberger et~al.(2009)Weinberger, Aubry, Lee, Nix, \protect\BIBand{}
  Cai}]{Weinberger09}
Weinberger, C.R., Aubry, S., Lee, S.W., Nix, W.D., Cai, W., 2009.
\newblock Modelling dislocations in a free-standing thin film.
\newblock Modelling and Simulation in Materials Science and Engineering 17,
  075007.

\bibitem[{Weygand et~al.(2002)Weygand, Friedman, {Van}~der Giessen,
  \protect\BIBand{} Needleman}]{Weygand02}
Weygand, D., Friedman, L.H., {Van}~der Giessen, E., Needleman, A., 2002.
\newblock {Aspects of boundary-value problem solutions with three-dimensional
  dislocation dynamics}.
\newblock Modell. Simul. Mater. Sci. Eng. 10, 437--468.

\bibitem[{Z-set(2020)}]{Zset20}
Z-set, 2020.
\newblock {Non-linear material \& structure analysis suite Z-set 9.1}.
\newblock {Mines ParisTech and French Aerospace Lab (ONERA)}, {Paris, France}.

\bibitem[{Zbib et~al.(1998)Zbib, Rhee, \protect\BIBand{} Hirth}]{Zbib98}
Zbib, H., Rhee, M., Hirth, J.P., 1998.
\newblock {On plastic deformation and the dynamics of 3D dislocations}.
\newblock Int. J. Mech. Sci. 40, 113--127.

\end{thebibliography}

\end{document}